\documentclass[journal]{IEEEtran}
\usepackage{cite}
\usepackage{mciteplus}
\usepackage{graphicx}
\usepackage{bm}
\usepackage{xcolor}

\usepackage{caption}
\usepackage{breqn}
\usepackage{xspace} 
\usepackage{algorithm,algorithmic}

\usepackage{cellspace}
\usepackage{tabularx,booktabs,caption,ragged2e}
\usepackage{xfrac}
\usepackage{amsmath}
\usepackage{amssymb}
\usepackage{array}
\usepackage{algorithm}
\usepackage{siunitx}
\usepackage{lipsum}
\usepackage[caption=false,font=normalsize,labelfont=sf,textfont=sf]{subfig}
\usepackage{epstopdf}
\begin{document}

\title{A Novel Approach to Secure RSMA Networks}

\author{Shaima Abidrabbu, and H\"{u}seyin Arslan, {\it Fellow, IEEE}
\thanks{S. Abidrabbu, and H. Arslan, are with the School of Engineering and Natural Science, Istanbul Medipol University, 34810 Istanbul, Turkey. (e-mail: shaima.abidrabbu@std.medipol.edu.tr, huseyinarslan@medipol.edu.tr).}
\thanks{S. Abidrabbu, is with IPR and License Agreements Department, Vestel Electronics, 45030 Manisa, Turkey. (e-mail: shaima.abidrabbu@vestel.com.tr).} \protect}

\maketitle

\begin{abstract}
This letter introduces a novel data-dependent interleaving technique designed to enhance the security of rate-splitting multiple access (RSMA) networks by protecting the common stream from eavesdropping threats. Specifically, we exploit the RSMA structure by interleaving the common bits of each user based on a sequence derived from their private bits. By decoding its private stream, the legitimate receiver reconstructs the interleaving sequence set by the transmitter and successfully de-interleaves the common stream. Therefore, the common part is successfully prevented from being intercepted by an eavesdropper who is unable to deduce the dynamic changing interleaving permutations. To ensure dynamic interleaving sequences, a private bit selection approach that balances the trade-off between security and system efficiency is proposed.  Simulation findings confirm the effectiveness of the suggested method, showing notable security improvements while maintaining robust overall system reliability.
\end{abstract}

\begin{IEEEkeywords}
RSMA, common stream, private stream, physical layer security, data-dependent interleaver.  
% \textcolor{red}{make sure that when we are using bits, or message, or stream, by looking to one of the published letter.}
\end{IEEEkeywords}

\section{Introduction}
Rate-splitting multiple access (RSMA) has recently emerged as a transformative non-orthogonal transmission technique and interference management strategy in wireless communications. It addresses critical limitations of traditional schemes such as non-orthogonal multiple access (NOMA) and space-division multiple access (SDMA) \cite{10315133}. By enabling robust performance under imperfect channel state information at the transmitter (CSIT), user mobility, and diverse operational conditions, RSMA enhances spectral efficiency, energy efficiency, user fairness, reliability, and quality of service \cite{ mao2022rate}. However, the broadcast nature of wireless communication introduces significant security vulnerabilities, rendering RSMA susceptible to both internal and external eavesdropping threats \cite{arslan2023physical}. Specifically, internal privacy vulnerabilities arise from RSMA's distinctive architecture, where the common stream—formed by aggregating data from all users—is broadcast to all users at higher power, making it more susceptible to internal eavesdropping. In contrast, private streams benefit from the relative security of being transmitted via channel-based SDMA precoding. This paper tackles the security concerns associated with the common stream by exploiting RSMA's inherent structural features to strengthen its defense against internal eavesdropping attacks. 

The integration of RSMA into secure communication systems
has demonstrated significant potential in optimizing security challenges across diverse wireless networks. For instance, Li et al. \cite{li2022rate} analyzed secrecy outage probabilities in satellite-terrestrial networks, highlighting RSMA’s resilience against eavesdropping and interference. Gao et al. \cite{gao2022rate} and He et al. \cite{he2023secure} offered secure RSMA designs by integrating intelligent reflecting surfaces (IRS) and addressing ergodic rates in land mobile satellite systems, respectively. Lu et al. \cite{lu2021worst} extended RSMA applications to energy-efficient secure communication in simultaneous wireless information and power transfer networks. On the other hand, several studies underscore RSMA’s versatility in enhancing and optimizing security through innovative strategies like beamforming, artificial noise, IRS integration, and cooperative relaying. For example, Cai et al. \cite{cai2021resource} and Xia et al. \cite{xia2022weighted} proposed secure RSMA beamforming and resource allocation schemes, maximizing secrecy rates under practical constraints. Nguyen et al. \cite{nguyen2023jamming} introduced a jamming-based covert communication technique for RSMA, enhancing secrecy while preventing detection. Cooperative rate-splitting (CRS) approaches by Zhang et al. \cite{zhang2019cooperative} and Li et al. \cite{li2020cooperative} utilized legitimate relays to confuse eavesdroppers and optimize secure sum rates. 
\begin{figure*}[h]
      \centering
        \includegraphics[scale=0.59]{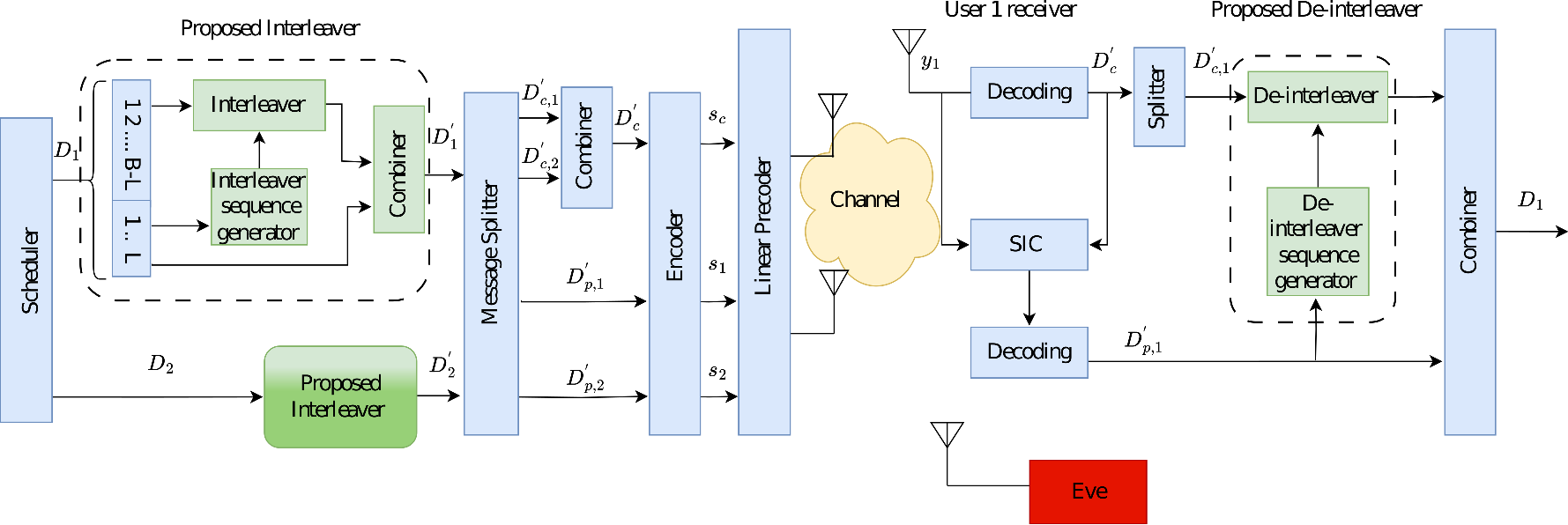}
        \caption{Proposed system model from transmitter and receiver perspectives.}
        \label{Fig2}
\end{figure*}

The aforementioned techniques rely on optimizing the overall system performance to enhance the security in RSMA networks. However, these optimization-based techniques introduce severe computational complexity to the system limiting their practicality for real-time applications. In this letter, we leverage the inherent structural properties of the RSMA to develop a novel framework that enhances resistance to eavesdropping threats, eliminating the reliance on intensive optimization techniques. The contributions of this letter are as follows:
\begin{itemize}
    \item A novel data-dependent interleaving approach that uses the intrinsic structure of RSMA to protect the common component of RSMA from both internal and external eavesdroppers is suggested. By linking the interleaving process to user-specific private data, the scheme makes patterns independent and highly unpredictable across transmissions. This mechanism, combining SDMA's spatial filtering with RSMA's dynamic interleaving, mitigates brute-force eavesdropping and enhances the overall security of RSMA networks significantly. 
    \item Moreover, a novel private bit selection strategy is proposed to achieve an adaptive trade-off between security and system efficiency. This approach effectively balances transmission reliability and eavesdropping resilience, ensuring robust and secure RSMA.
    \item The proposed approach is derived and tested in terms of bit error rate (BER). Simulation results demonstrate that the proposed approach effectively enhances security without compromising spectral efficiency, highlighting its practical viability for secure RSMA networks.
\end{itemize}

\section{SYSTEM MODEL}
Consider a downlink cellular network with a single base station (BS) equipped with $N_t$ transmit antennas, serving  $K$ legitimate single-antenna users. The system employs RSMA to enhance security in the presence of a passive, silent eavesdropper capable of demodulating RSMA signals. This eavesdropper can intercept all transmissions between the BS and legitimate users but does not interfere with legitimate communications. In the RSMA framework, each user’s message $D_k$ is divided into a common part $D^{'}_{\mathrm{c}, k}$ and a private part $D^{'}_{\mathrm{p},k}$. The common parts across all users $\left\{D^{'}_{c, 1}, \ldots, D^{'}_{c, K}\right\}$ are jointly encoded into a common stream $s_{\mathrm{c}}$, while the private parts $\left\{D^{'}_{p, 1}, \ldots, D^{'}_{\mathrm{p}, K}\right\}$ are encoded separately into private streams $\left\{s_1, \ldots, s_K\right\}$. The combined data streams are represented as $\mathbf{s}=\left[s_{\mathrm{c}}, s_1, \ldots, s_K\right]^T \in \mathbb{C}^{K+1}$, and undergo linear precoding before transmission using the precoding matrix $\mathbf{P}=\left[\mathbf{p}_{c}, \mathbf{p}_1, \ldots, \mathbf{p}_K\right]$, where $\mathbf{p}_{c}, \mathbf{p}_{k} \in \mathbb{C}^{N_t \times 1}$ are the precoders for the common and private streams, respectively. The BS utilizes the channel state information (CSI) of all users to derive the precoding matrix for transmission. Therefore, the signal transmitted by the Base station can be expressed as follows: 
\begin{equation}
    \textbf{s}=  {\textbf{p}_{c}} s_{\mathrm{c}} + \sum_{k=1}^{K} {\textbf{p}_{k}} s_{k}, \quad \forall k \in {K}. 
\end{equation}  

The received signal for the $k$-th user is expressed as:
\begin{equation}
  {y}_{k}=\mathbf{h}_{k}^{H} \textbf{s}+{n}_{k}, 
\end{equation}
where $\mathbf{h}_{k} \in \mathbb{C}^{N_{t} \times 1}$ denotes the channel gain from the BS to the $k$-th user, known perfectly at the BS, and ${n}_{k} \sim \mathcal{CN}\left(0, \sigma_{k}^{2}\right)$ represents the additive white Gaussian noise (AWGN) with zero mean and variance $\sigma_{k}^{2}$. To ensure power constraints, we assume $\mathbb{E}\left\{\textbf{ss}^H\right\}=\mathbf{I}$ and impose a total transmit power budget $P_{\mathrm{t}}$, such that $\operatorname{tr}\left(\mathbf{P} \mathbf{P}^H\right) \leq P_{\mathrm{t}}$. Each component of the splitted messages adhere to separate power limits: $P_{k}$ for private messages ($\sum_{k=1}^{K} p_{k} \leq P_{\mathrm{k}}$) and $P_{c}$ for common streams ($p_{c} \leq P_{\mathrm{c}}$) where $p_{k}$ and $p_{c}$ are the transmit powers of the private and common streams for the $k$-th user. Thus, the combined power satisfies $\sum (P_{\mathrm{k}}+P_{c})\leq P_{\mathrm{t}}$. The RSMA-enabled transmitter includes a “message splitter,” dividing each user’s message into $Z$ sub-messages, facilitating rate-splitting. The number of sub-messages $Z$ depends on the RSMA scheme employed, which can vary according to message composition, upper-layer protocols, and system design criteria. Existing literature predominantly utilizes a concatenated message-splitting approach.

\section{THE PROPOSED SECURE RSMA APPROACH}
\subsection{Definition of the Proposed Approach}
Figure \ref{Fig2} illustrates the proposed transmitter structure for an RSMA network, designed to enhance security against eavesdropping. In this framework, the scheduler processes the data for each user by dividing it into two segments: $B$ bits and $L$ bits. The $B$ bits are interleaved using a sequence generated by the interleaver, which leverages the $L$ bits to create the interleaving pattern. This process introduces an additional layer of security, increasing the resilience of the RSMA network by making it significantly more complex for internal eavesdroppers to reconstruct the common information belonging to other RSMA users. The proposed interleaver generates a shuffled version of the common part of each user, restructuring the data sequence before the conventional message-splitting stage in the RSMA network as 
\begin{equation}
  D^{'}_k=\{\underbrace{D^{'}_1, D^{'}_2, D^{'}_3, \dots, D^{'}_B}_{\text{Common part (shuffled)}}, \underbrace{D_1, D^{'}_2, D^{'}_3, \dots, D_L}_{\text{Private part (unshuffled)}}\}.  
  \label{eq3}
\end{equation} 

Subsequently, the message splitter processes the modified data of the $k$-th user by assigning all bits from the $L$ section to the private stream. These bits are secured through user-specific precoding implemented via the linear precoder, ensuring their confidentiality. This interleaving process significantly complicates eavesdroppers' reconstruction attempts, as the interleaving sequence dynamically changes for different data streams. The interleaved common streams are then combined and transmitted over the channel alongside the private streams. At the legitimate receiver, both the common and private streams are received with a power difference, as per the RSMA framework. The decoding process begins with the common stream, followed by successive interference cancellation (SIC) to decode the private stream, maintaining the conventional RSMA operation. However, prior to combining, the legitimate receiver reconstructs the interleaving sequence from the private stream and unshuffles the common data accordingly, ensuring accurate decoding and combining. In contrast, at the eavesdropper’s receiver, the shuffled common stream introduces additional uncertainty, significantly hindering attempts to decipher the transmitted information. The interleaving pattern is autonomously generated at both the transmitter and the legitimate receiver using the private data of the intended user, thereby eliminating the need for shared keys. The interleaving patterns are dynamically generated using the interleaver generator formula, which can be configured for exclusivity in critical applications.

\subsection{Addressing Bit Selection}
To address the challenges posed by the error-prone nature of wireless channels and the absence of perfect channel reciprocity, we propose interleaving a subset of $B$ bits from each RSMA private signal. This subset is chosen to strike a balance between transmission reliability and resistance to eavesdropping, even in the presence of channel variations. Let $\mathcal{F}$ denote the set of interleaved bits, with the selection for each bit represented by $\zeta(b)$ for each bit is given by:  
\begin{equation} 
\zeta(b) = \begin{cases} 1, & b \in \mathcal{F} 
\\ 0, & b \notin \mathcal{F} 
\end{cases} \quad b = 1, \ldots, B-1. 
\label{eq4}
\end{equation}
While data reciprocity enables the legitimate receiver to estimate $\zeta$ based on its private data detection, imperfections in the decoding may introduce slight deviations. The transmitter securely determines the interleaving selected bits $\zeta$, which is then identified by the legitimate receiver during the decoding process. In contrast, eavesdroppers encounter significant difficulty in reconstructing $\zeta$ without access to the private stream or the transmitter's initial configuration, forcing them to rely on random guessing or exhaustive brute-force attempts.

\subsection{Resistance to Brute-Force Attacks}
In brute-force attacks, an eavesdropper attempts to decode the interleaving pattern by testing all the possible permutations. The proposed RSMA scheme counters this by dynamically updating the interleaving pattern during each channel coherence interval, using the private data bits designated for transmission at that time. This dynamic process ensures interleaving patterns are independent across transmissions, making previously intercepted data useless for future decoding. The continual updates and random generation of patterns based on user-specific private bits add substantial complexity, significantly
hindering the eavesdropper’s ability to reconstruct the pattern and decipher the interleaved data.

\subsection{Interleaving Permutation Pattern}  
The proposed interleaving mechanism generates interleaved bits based on a set of private bits, denoted as $\mathcal{L}_{i}$, which serve as indexing bits. These bits play a critical role in defining the data interleaving patterns for the $k$-th user. The number of interleaving sequences is determined by $S_{\text{In}} \leq \min \left(2^{\mathcal{L}_{i}}, B!\right)$, where $2^{\mathcal{L}_{i}}$ represents the sequences defined by indexing bits, and $B!$ denotes the total permutation space of the interleaving sequence.
Each user’s common data stream is a function of the interleaved bits, expressed as $B=\mathfrak{f}(\mathcal{L}_{i})$. The equation $B + L= D_k^{'}$ ensures that each user’s data consists of both private (unshuffled) and common (shuffled) components. Not all private bits are required for interleaving, resulting in three distinct components in the $k$-th user’s data:
\begin{enumerate}
    \item Interleaved bits (common stream).
    \item Indexed private bits (used for generating the interleaving sequence).
    \item Non-indexed private bits (unused in interleaving generation).
\end{enumerate}
The distribution of bits can be expressed as:
\begin{equation} 
D_u^{'} = D_k^{'} - (B + L) = D_k^{'} - L - \mathfrak{f}(\mathcal{L}_{i}),  \label{44} 
\end{equation}
where $D_u^{'}$ represents the non-indexed private bits. Consequently, the private bits are given by $D_{p,k} = L + D_u^{'}$, and the ratio of uninterleaved to total bits is:
\begin{equation} 
\rho = \frac{D^{'}_{p,k}}{D_k^{'}} = \frac{L + D_u^{'}}{L + D_u^{'} + B}. 
\end{equation}

To analyze the proposed interleaving mechanism, an algorithm is outlined to map indexing bits into a corresponding bit interleaving sequence. The generation of the interleaving sequence begins with setting $\mathcal{L}_{i}=B-1$, implying that $\mathfrak{f}(\mathcal{L}_{i})=\mathcal{L}_{i}+1$. Consequentially, $S_{\text{In}}$ is calculated as $2^{\mathcal{L}_{i}-1}$. From a physical layer security perspective, minimizing the number of unscrambled bits (i.e, $D^{'}_u$) is crucial to maximize the potential sequence combinations, as demonstrated in Table \ref{title4}. As an example, the interleaving sequence $\textbf{\mathcal{Q}}$ is constructed using the specified parameters and follows the principles of Gray mapping. The algorithm initiates with an identity interleaving sequence $\textbf{\mathcal{Q}}$. For each indexing bit, specific bits of $\textbf{\mathcal{Q}}$ are exchanged depending on the bit’s value. For example, if the first indexing bit equals one, bits 1 and 2 are swapped; otherwise, the interleaving sequence remains unchanged. The process continues sequentially: if the second indexing bit is one, bits 2 and 3 are swapped; otherwise, no changes are made. This pattern repeats for all remaining indexing bits, with the corresponding bits being adjusted accordingly. The design ensures that if two indexing sequences differ by only a single bit, their respective interleaving sequences will differ in just one pair of exchanged bits in $\textbf{\mathcal{Q}}$. This property minimizes the impact of errors in the indexing bits on the overall BER, thereby enhancing the system’s resilience to bit errors. The overall algorithm is represented in Algorithm \ref{aLGORITHM1}. 

The computational complexity of the proposed approach differs for legitimate users and eavesdroppers. For legitimate users, the complexity is $\mathcal{O}(B/2)$, determined by the known value of $B$ and the implementation of Gray mapping for pattern generation. Conversely, eavesdroppers encounter substantially higher complexity, quantified as $\mathcal{O}(B!/(B-\mathcal{L}_{i})!)$, due to the need to account for all possible interleaving sequences without prior knowledge of $B$ and $\mathcal{L}_{i}$.  
\begin{table}[!h]
 \caption {Design parameters for different configurations.}
\label{title4}
\renewcommand{\arraystretch}{1.42}
\resizebox{1\columnwidth}{!}{
\begin{tabular}{|l|l|l|l|l|l|l|}
\hline
$B$ & $L$ & $b$ & $\mathcal{L}_{i}$ & $S_{\text{In}}$ & $D_u^{'}$ & $\rho$ (\%) \\ \hline
$50$  & $50$  & $25$  & $25$ & $16777216$        & $25$        & $60\%$        \\ \hline
$37$  & $63$  & $15$  & $63$  & $4.6 \times \num{e18}$ & $0$         & $63\%$        \\ \hline
$37$  & $63$  & $37$  & $30$  & $536870912$       & $33$        & $72\% $       \\ \hline
$37 $ & $63$  & $15$  & $15$  & $16384$           & $48$        & $75\% $       \\ \hline
\end{tabular}}
\end{table}

\begin{algorithm}
\begin{algorithmic}[1]
 \STATE  \textbf{Input} : $B-1 \times 1$ vector of indexing bits $\mathcal{L}_{i}$. 
 \STATE  \textbf{Output}: Interleaving sequence  $\textbf{\mathcal{Q}}$ of size $B \times 1$. 
 % \STATE   $\textbf{\mathcal{Q}} \leftarrow \mathbf{I}_{B \times B}$.
    \FOR{$b \leftarrow 1$ to $B-1$} 
         \STATE $f \leftarrow b+1$
        \IF{$\mathcal{L}_{i}(b)=1$} 
            \STATE Exchange $b$-th and $f$-th bits of $\textbf{\mathcal{Q}}$
        \ELSE
            \STATE Do nothing
        \ENDIF
    \ENDFOR
\caption{Interleaving pattern generation.}
\label{aLGORITHM1}
\end{algorithmic}
\end{algorithm}

\section{PERFORMANCE EVALUATION} 
In this section, the BER for the $k$-th user is analyzed  under Gray coding. Assumed that $M$-phase-shift keying (PSK) is utilized where the constellation size is $M=2^v$ corresponds to $v$ bits per symbol, the BER of the $k$-th user is given by 
\begin{equation}  
BER_{k}=\rho BER^{\mathrm{p}}+(1-\rho) BER^{\mathrm{c}},  
\end{equation}
where $BER^{\mathrm{p}}$, and $BER^{\mathrm{c}}$ denote the BEPs for the private and common streams, respectively. The $BER^{\mathrm{p}}$ can be calculated as
\begin{equation}
BER^{\mathrm{p}}=\frac{1}{D_u^{'}} \sum_{l=1}^L Pr (\gamma_l),  
\end{equation}
where $Pr (\gamma_l)$ represents the BEP of the $l$-th private bits at the $l$-th SNR $\gamma_l$. For the AWGN channel, the theoretical BEP can be defined as \cite{1021039} 
\begin{equation}
Pr (\gamma_l)=\sum_{m=1}^{(B+L)} \frac{1}{(B+L) \sqrt{M}} \sum_{l=0}^{\left(1-2^{-k}\right) \sqrt{M}-1} \Gamma_{m,l}~,  
\end{equation}
where $(B+L)$ is the total number of bits for the $k$-th user and $\Gamma_{m,l}$ is the Gamaa function given as
\begin{equation}
\begin{aligned}
\Gamma_{m,l} & =(-1)^{\left\lfloor\frac{l 2^{m-1}}{\sqrt{M}}\right\rfloor}\left(2^{m-1}-\left\lfloor\frac{l 2^{m-1}}{\sqrt{M}}+\frac{1}{2}\right\rfloor\right) \\
& \times \operatorname{erfc}\left((2 l+1) \sqrt{\frac{3 \log _2(M) \gamma_l}{2(M-1)}}\right),
\end{aligned}  
\end{equation}
where the $\operatorname{erfc}(x)$ is the complementary error function. On the other hand, $BER^{\mathrm{c}}$ is expressed as 
\begin{equation}
\begin{aligned}
BER^{\mathrm{c}} & =\operatorname{Pr}\{q=0\} \sum_{b=1}^{B} \frac{Pr(\gamma_b)}{B} \\
& +\sum_{l=1}^{L} \operatorname{Pr}\{q=l\} \sum_{b=1}^{B} \frac{Pr(\gamma_b \mid q=l)}{B},
\end{aligned}  
\end{equation}
where $\operatorname{Pr}\{q=l\}$ denotes the probability of $l$ bits in error 
among $L$ indexing bits, and $Pr(\gamma_b \mid q=l)$ is the BEP of the $M$-QAM symbol on $b$-th bit given $q$-bit-error in the indexing bits (i.e, $q=0, 1, 2, ..., \mathcal{L}_{i}$). At high SNR, the most probable error events for Gray-coded involve single-bit errors, allowing an approximation where an $q$-bit error event in the indexing bits induces $q$ simultaneous bit errors on the indexing bits. When SNR is moderate to high, with indexing bits significantly exceeding the number of erroneous bits in $L$, an interleaving sequence design where $l$ bit errors alter the positions of $(l+1)$ QAM symbols upon deinterleaving yields the approximation:
\begin{equation}
Pr(\gamma_b \mid q=l)\cong\left(\frac{B-l-1}{B}\right) Pr(\gamma_b)+\frac{1}{2}\left(\frac{l+1}{B}\right) .   
\end{equation}
The probability of $l$ erroneous indexing bits can be giving as 
\begin{equation}
Pr\{q=l\}=\binom{L}{l}\left\{Pr(\gamma)\right\}^l\left\{1-Pr(\gamma)\right\}^{L-l},  
\end{equation}
where $\binom{L}{l}$ denotes the binomial coefficient. Assuming a frequency-flat channel where all bits have identical average SNR $\bar{\gamma}$, as $\bar{\gamma}$ increases, the probability of a single erroneous bits after deinterleaving dominates, allowing the overall simplification as follows 
\begin{equation}
\begin{aligned}
BER_{k} & \approx \frac{\left(D_u+B ~ Pr\{q=0\}\right) Pr(\bar{\gamma})}{B+L} \\
& +\frac{\left(\left(B-2\right) Pr(\bar{\gamma})+1\right) Pr\{q=1\}}{B+L}.
\end{aligned} 
\end{equation}

\section{SIMULATION RESULTS}
This section displays the BEP curves for the legitimate user while utilizing the proposed interleaving procedure and investigates the error gap observed in comparison to both internal and external eavesdroppers and conventional networks. In all scenarios, the amount of total bits for a user is set to $100$, $B=25$, $L=50$, $\mathcal{L}_{i}=25$, $D_u^{'}=25$, the modulation technique quadrature PSK (QPSK) is used, and the SNR is changed from -10 to 20 dB in 1 dB steps. 

\begin{figure*}[!t]
\centering
\subfloat[]{\includegraphics[width=3.5in]{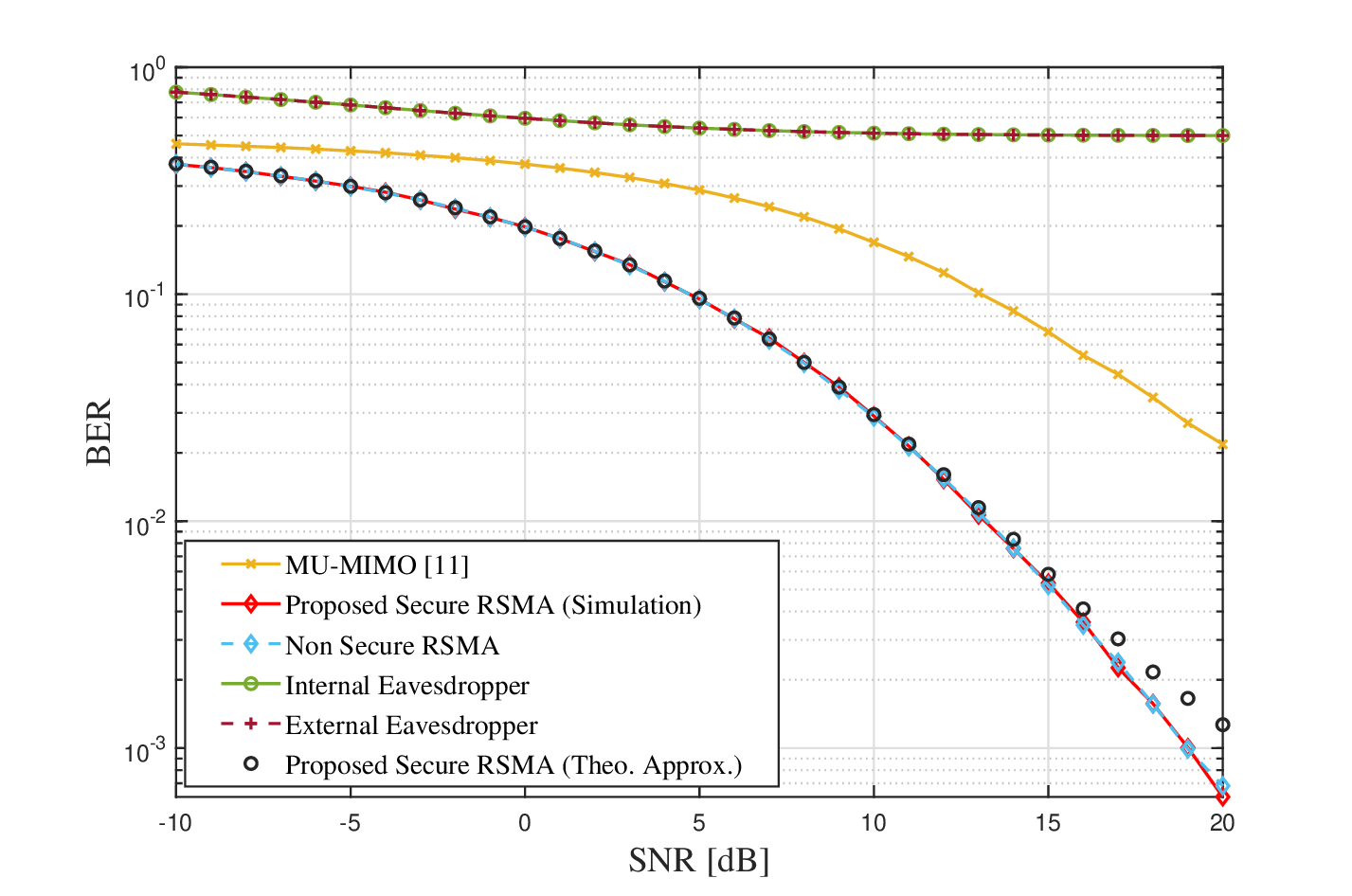}%
\label{fig22}}
\hfil
\subfloat[]{\includegraphics[width=3.5in]{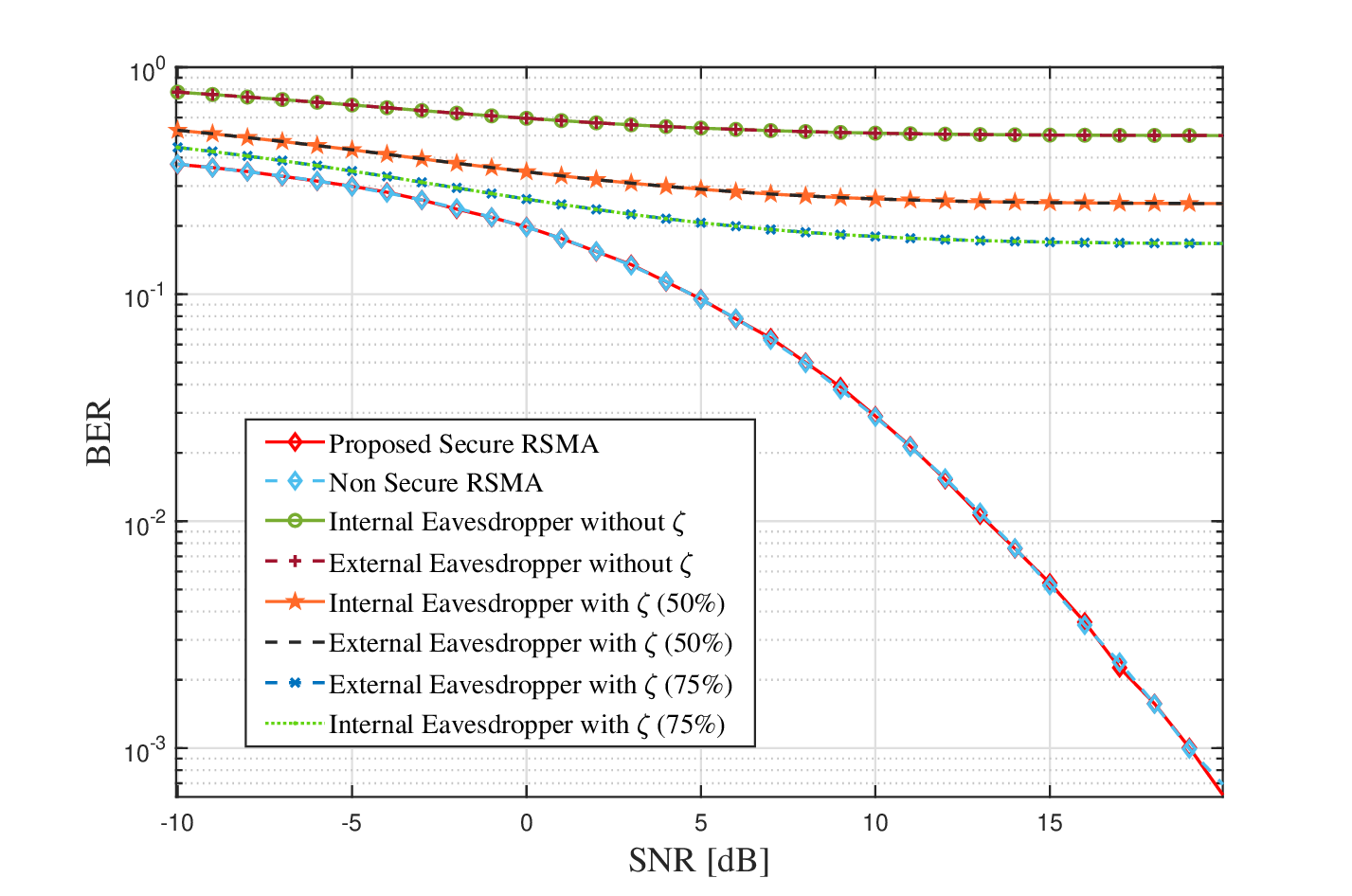}%
\label{fig33}}
\caption{ BER evaluation: (a) the proposed secure RSMA approach, and (b) the proposed secure RSMA  with and without the side information $\zeta$.}
\label{fig_sim}
\end{figure*}

Figure \ref{fig_sim}(a) presents the BER performance comparison between the proposed secure RSMA scheme, a non-secure RSMA configuration, and multi-user MIMO (MU-MIMO) \cite{cho2010mimo}. The results highlight the proposed method's superior security performance against both internal and external eavesdropping threats while maintaining the integrity of normal RSMA communication. This enhancement is achieved by leveraging the inherent private data of each user to secure the common stream, demonstrating a novel application of existing RSMA network features rather than introducing additional overhead. Notably, the proposed approach preserves the data rate and spectral efficiency of standard RSMA communication, ensuring no degradation in overall system performance.

The effectiveness of the proposed approach in preventing eavesdropping is evaluated by analyzing the impact of side information $\zeta$ on the eavesdropper's ability to decode the common stream. Fig. \ref{fig_sim}(b) compares BER performance for eavesdroppers with and without knowledge of a percentage of $\zeta$. In the simulations, an eavesdropper lacking any side information ($\zeta$) attempts to estimate the bit selection using a shuffled version of the common signal. However, even with such estimations, the eavesdropper is unable to decode the common stream, resulting in significantly higher BERs compared to the legitimate user, who has precise knowledge of the interleaved sequence. The analysis further assumes that the eavesdropper has partial access to $\zeta$, specifically 50\% and 75\%. The results show that even with 75\% knowledge of $\zeta$, the eavesdropper's BER remains high, underscoring the robustness of the proposed method. These findings demonstrate the approach's strong resilience against both internal and external eavesdropping attempts, effectively securing the RSMA network.

\section{CONCLUSION AND FUTURE WORK}
This letter proposed a novel data-dependent interleaving mechanism aimed at enhancing the security of RSMA networks by utilizing the private data of legitimate users. The approach dynamically generates interleaving sequences, significantly increasing the complexity for internal and external eavesdroppers attempting to intercept the common stream. The proposed scheme’s efficacy was validated through simulation results and a closed-form approximation of the BEP at the legitimate receiver. The results demonstrate the method's ability to secure RSMA networks effectively without degrading spectral efficiency or communication performance. Future work may explore adaptive bit selection strategies that dynamically adjust secrecy levels based on individual user requirements. Such advancements could further optimize network performance by achieving an improved trade-off between security and system efficiency, ensuring adaptability to a variety of application scenarios.

\bibliographystyle{IEEEtran}
\bibliography{IEEEabrv.bib,ref.bib}{}
\end{document}